\documentstyle[ltwol]{article}

\input{psfig}

\def\Journal#1#2#3#4{{#1} {\bf #2}, #3 (#4)}

\def\ICFA{\em ICFA Instr. Bullettin}

\def\NIM{\em Nucl. Instrum. Methods}

\def\etal{{\em et al.}~}
\def\epm{$e^+e^-$}
\def\ufs{$\Upsilon(4S)$}

\def\be{\begin{equation}}
\def\ee{\end{equation}}
\def\bea{\begin{eqnarray}}
\def\eea{\end{eqnarray}}

\begin{document}

\vskip 1.0cm
\twocolumn[
\rightline{\vbox{\hbox{HEPSY 98-2}
\hbox{October, 1998}}}
\vskip 1.0cm
\centerline {\bf PROGRESS TOWARDS CLEO III}
\centerline{\bf THE SILICON TRACKER AND THE
LIF-TEA RING IMAGING CHERENKOV DETECTOR}
\centerline{\it Talk
 presented at the
XXIX International Conference on High Energy Physics  }
\centerline{\it Vancouver, Canada, July 1998}
\normalsize
 \vskip 1.0cm
\centerline{Marina Artuso}
\centerline{Department of Physics,}
\centerline{Syracuse University,}
\centerline{Syracuse, New York 13244--1130}
\centerline{\it e-mail: artuso@physics.syr.edu}
\vskip 1.0cm
 

\abstracts
{We describe the two major components of CLEO 
III:
the Silicon Vertex 
Detector and the Ring Imaging
 Cherenkov Detector (RICH).  The Silicon Vertex Detector is a four layer
 barrel-style device
which spans the radial distance from 2.5 cm to 10.1 cm and covers 93\%
of the solid angle. It is being constructed
using double-sided silicon sensors read out by front
end electronics devices especially designed for this application.
The RICH system
consists
of LiF radiators and multiwire proportional chambers containing a mixture of
CH4 and TEA gases. The radiators are both flat and "sawtooth."
Results from a test beam run of final CLEO III RICH modules will be
presented, as well as test beam data on sensors to be employed in the Silicon 
Vertex Tracker.}]

\vskip 1.0cm
\section{Introduction}
The Cornell \epm\ collider (CESR) is currently being upgraded to a luminosity 
in excess of 
$1.7\times 10 ^{33} {\rm cm }^{-2}{\rm s}^{-1}$. In parallel the CLEO III 
detector is undergoing some major improvements. 

One key 
element of this upgrade is the construction of a four layer double-sided
silicon tracker. 
This detector
spans the radial distance from 2.5 cm to 10.1 cm and cover 93\%
of the solid angle surrounding the interaction region.  The
outermost layer is 55 cm long and will present a large
capacitive load to the front-end electronics. The innermost layer must
be capable of sustaining large singles rates typical of a detector
situated near an interaction region.

A novel feature of CLEO III is a state of the art particle identification system that
will provide excellent hadron identification at all the momenta relevant to 
the study of the decays of B mesons produced at the \ufs\ resonance. The technique
chosen is a proximity focused Ring Imaging Cherenkov detector 
(RICH)~\cite{upsil} 
in a barrel geometry occupying 20 cm of radial space between the tracking system
and the CsI electromagnetic calorimeter.

The physics reach of CLEO III is quite exciting: the increased sensitivity of the upgraded
detector, coupled with the higher data sample available, will provide a great sensitivity 
to a wide variety of rare decays, CP violating asymmetries in rare decays and precision 
measurements of several Cabibbo-Kobayashi-Maskawa matrix elements.

\section{Vertex Detector Design}
The barrel-shaped CLEO III Silicon Tracker (Si3) is composed of 447 identical sensors combined 
into 61 ladders. The sensors are double sided with ${\rm r}\phi$ readout on the n side and 
z strips on the p side. The strip pitch is 50 $\mu$m on the n side and 100
$\mu$m on the p side. 
Readout hybrids are attached at both ends of the ladders, each reading
out half of the ladder sensors. More details on the detector design can be found elsewhere.~\cite{ian} Sensors and 
front end electronics are connected by flex circuits that have traces with a 100 $\mu$m pitch on 
both sides, being manufactured by General Electric, Schenectady, New York. 

All the layers are composed of identical sensors. In order to simplify the sensor 
design, the detector biasing resistors and the coupling capacitors have been removed from 
the sensor into a dedicated R/C chip, mounted on the hybrid. Another key feature in the sensor 
design is the so called ``atoll'' geometry of the p-stop barriers, using isolated p-stop rings
surrounding individual n-strips. Furthermore a reverse bias can be applied to the p-stop barriers through a
separate electrode. Thus 
the parasitic capacitance associated with these insulation barriers can be significantly
 reduced with a corresponding reduction of the sensor noise in the frequency range 
of interest.~\cite{george}
  
The middle chip in the readout chain is the FEMME preamplifier/shaper VLSI device.
It has an excellent noise performance. At the shaping time of 2$\mu$s, well matched to the CLEO III
trigger decision time, its equivalent noise charge is measured as:
\begin{equation}
ENC = 149 e+ 5.5 e/pF \times C_{in},
\end{equation}
giving satisfactory noise performance also with the high input capacitances in the outer layer
sensors. More details on the design and performance of this device can be found elsewhere.~\cite{osu}
The last chip in the readout chain is the SVX\_CLEO digitizer and 
sparsifier.~\cite{george} Both these chips are manufactured utilizing radiation -hard CMOS technology from 
Honeywell. 

\section{The Si3 test beam run}
The silicon sensors have been tested in several test beam runs that took place at CERN in the
last few months. The sensors, flex circuits, and R/C chips used were the same ones planned
for the final system. However the readout electronics used was not
the combination of FEMME + SVX\_CLEO, but the low noise VA2 chip produced
by IDE AS, Norway~\cite{einar} with digitization implemented in the remote data acquisition 
system. 

The data were collected inserting the Si3 sensor in the test beam set-up used by the RD42
collaboration~\cite{rd42} to test their diamond sensors.  A 100 GeV $\pi$ beam was used and the
silicon sensor was inserted in a silicon telescope composed of 8 microstrip reference planes
defining the track impact parameters with a precision of about 2 $\mu$m. Two data sets were
collected. The first one contains 300,000 events with tracks at normal incidence and different
bias points. The second one consists of about 200,000 events at $\theta =0$ 
and 
$\theta =10 ^{\circ}$. 
\begin{figure}
\begin{center}

\vspace{-1.4cm}
\psfig{figure=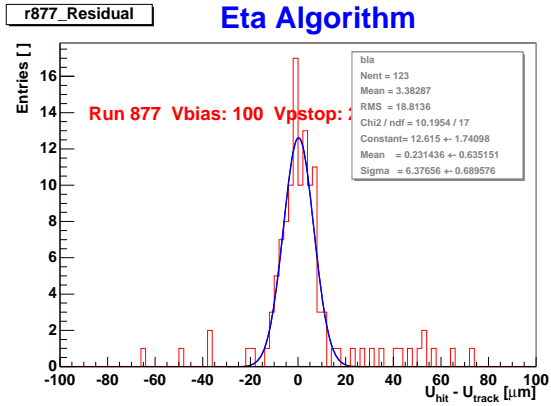,height= 3.5in}
\vspace{-1.7cm}
\caption{Residual distribution with $\eta$ algorithm at $V_{bias}$=100 V and
$V_{pstop}=20 V$.}
\end{center}
\label{fig:resol}
\end{figure}
 The probability distribution of the variable $\eta$
defined as $\eta =Q_L/(Q_L+Q_R)$ (referring to the relative location of
two adjacent strips where at least one of them recorded a hit) has 
been mapped and has been used in a non-linear charge 
weighting to reconstruct the track hit location.~\cite{weil}  Fig.~1 shows
the hit resolution achieved on the n-side with a detector bias of 100 V and
a p-stop reverse bias of 20 V. A value of $\sigma$ of 6 $\mu$m is quite impressive for a track
at normal incidence. The expectation for a 50 $\mu$m strip pitch is about
 9 $\mu$m, as the charge spreading due to diffusion does 
not provide appreciable signal in the neighboring strip unless the
track impact point is at the periphery of the strip. The higher resolution
achieved in this case is attributed to an increase in charge spread due to the
reverse biased p-stops. In fact the ability to modulate the reverse bias of the
p-stops has allowed the tuning of this charge sharing for optimum 
resolution and may be of interest for other applications.

\section{Description of the CLEO III RICH System}
The CLEO III RICH system is based on the `proximity focusing' approach, in 
which 
the Cherenkov cone, produced by relativistic particles crossing a LiF radiator,
 is let to expand in a volume filled with gas transparent to
ultraviolet light before intersecting a photosensitive detector where the 
coordinates of the Cherenkov photons are reconstructed. The photodetector is amultiwire proportional chamber filled with a mixture of Triethylamine (TEA) gas 
and Methane. The TEA molecule has good quantum efficiency (up to 35\%) in
a narrow wavelength interval between 135 and 165 nm and an absorption length of
only 0.5 mm. 

The position of the photoelectron emitted by the TEA molecule
upon absorption of the Cherenkov photon is detected by sensing the induced 
charge on an array of $7.6{\rm mm} \times 8 $ mm cathode pads. The probability distribution for the charge in the avalanche initiated by a single photoelectron
is exponential at low gain. This feature implies that a low noise 
front end electronics is crucial to achieve good efficiency. A dedicated 
VLSI chip, called VA\_RICH, based on a very successful chip developed for 
solid state application, has been designed and produced for our application
at IDE AS, Norway. We have acquired and characterized all the hybrids necessary
to instrument the whole RICH detector, a total of 1800 hybrids containing 
two VA\_RICH chips each, corresponding
230,400 readout channels. We have fully characterized all of them and for 
moderate values of the input capacitance $C_{in}$, the equivalent noise 
charge $ENC$ is found to be about:
\begin{equation}
ENC = 130 e^- + (9e^- /pF)\times C_{in}.
\end{equation}
The traces that connect the cathode pads with the input of the preamplifier in
the VA\_rich are rather long and the expected value of $ENC$ in absence of 
other contribution is of the order of 200 $e^-$.

The charge signal is transformed into a differential current output transmitted
serially by each hybrid to a remote data acquisition board, where the currents
are transformed into voltages by transimpedance amplifiers and then digitized by
a 12 bit flash-ADC capable of digitizing the voltage difference at its input.
The data boards perform several additional complex functions, like providing
the power supply and bias currents necessary for the VA\_RICH to be at its optima
working point. In addition, the digital component of these boards provides 
sparsification, buffering and memory for pedestal and threshold values.

If a track crosses the LiF radiator at normal incidence, no light is emitted 
in the wavelength range detected by TEA, due to total internal reflection. 
In order to overcome this problem a novel radiator geometry has been 
proposed.~\cite{alex} It involves cutting the outer surface of the radiator like the
teeth of a saw, and therefore is referred to as ``sawtooth radiator''. A detailed
simulation of several possible tooth geometries has been performed and a
tooth angle of 42$^{\circ}$ was found out to be close to optimal and technically
feasible.~\cite{alex}

There are several technical challenges in producing these radiators, including
the ability of cutting the teeth with high precision without cleaving the
material and polishing this complex surface to yield good transmission properties
for the ultraviolet light. One of the goals of the test beam run described below 
was to measure the performance of sawtooth radiators and we were able to produce 
two full size pieces working with OPTOVAC in North Brookfield, Mass. The 
light transmission properties of these two pieces were measured relative to a 
plane polished sample of LiF and found to be very good.~\cite{icfa}

\section{Test beam results} 
Two completed CLEO III RICH modules were taken to Fermilab and exposed 
to high energy muons emerging from a beam dump. Their momentum was $\ge$ 100 GeV/c
The modules were mounted on a leak tight aluminum box with the same mounting 
scheme planned for the modules in the final RICH barrel. One plane radiator and
the two sawtooth radiators were mounted inside the box at a distance from the
photodetectors equal to the one expected in the final system. Two sets of multiwire
chambers were defining the $\mu$ track parameters and the trigger was 
provided by an array of scintillator counters. The data acquisition system was
a prototype for the final CLEO system.

The beam conditions were much worse than expected: the background 
and particle fluxes were about two order of magnitude higher than we expect
in CLEO and in addition included a significant neutron component that is going
to be absent in CLEO. Data were taken corresponding to different track incidence
angles and with tracks illuminating the three different radiators. For the
plane radiator, we were able to configure the detector so that the photon
pattern would appear only in one chamber. For the sawtooth radiator a 
minimum of three chambers would have been necessary to have full acceptance.

The study of this extensive data sample has been quite laborious and the full
set of results is beyond the scope of this paper. The results from two typical
runs will be summarized in order to illustrate the expected performance from our system:
the first case will involve tracks incident at 30$^{\circ}$  to the plane
radiator and the second tracks incident at 0$^{\circ}$ on a sawtooth radiator.

\begin{figure}
\begin{center}
\vskip 0.5cm
\psfig{figure=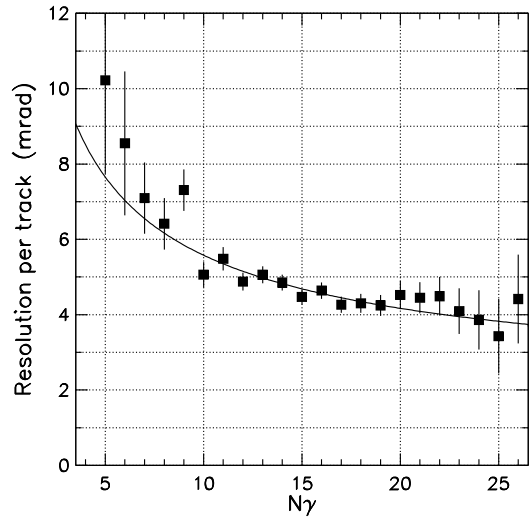,height= 3.5 in}
\vspace{-1.0cm}
\caption{The Cherenkov angular resolution per track as a function of the number
of detected photons (background subtracted) for a plane radiator with tracks
at 30$^{^\circ}$ incidence.}
\end{center}
\label{fig:sngf}
\end{figure}

\begin{figure}
\begin{center}
\vspace{0.8cm}
\psfig{figure=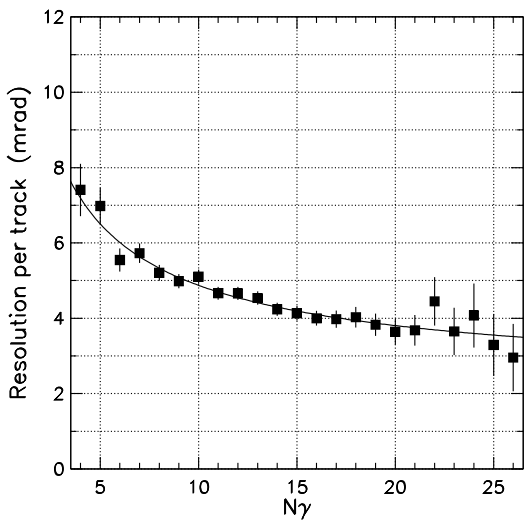,height= 3.5 in}
\vspace{-2.0cm}
\caption{The Cherenkov angular resolution per track as a function of the number
of detected photons (background subtracted) for a sawtooth radiator with tracks
at normal incidence.}
\end{center}
\label{fig:sngs}
\end{figure}

The fundamental quantities that we study to ascertain the expected performance
of the system are the number of photons detected in each event and the angular
resolution per photon. The measured distributions are compared with the predictions of
a detailed Monte Carlo simulation, including information on the CH$_4$-TEA 
quantum efficiency as a function of wavelength, ray tracing, crystal transmission, etc.. It includes also a rudimentary model of the background, attributed 
to out of time tracks. The agreement between Monte Carlo
and data is quite good. The average number of detected photoelectrons is about
14 after background subtraction and the angular resolution per photon is
about 13.5$\pm$ 0.2 mr.
  The average number of photons, after background subtraction, is
13.5, with a geometrical acceptance of only 55\% of the final system. In this
case it can be seen that the expected resolution is slightly  better than 
the one achieved. In order to estimate the particle identification power in our system,
we need to combine the information provided by all the Cherenkov photons in
 an 
event. We can use the resolution on the mean Cherenkov angle per track as an estimator
for the resolving power in the final system. 

\begin{table}
\begin{center}
\caption{Summary on the performance of the RICH modules in two typical 
test beam runs. The symbol $\gamma$ refers to single photon 
distributions and the symbol $t$ refers to quantities averaged
over the photons associated with a track.}\label{tab:smtab}
\vspace{0.5cm}
\begin{tabular}{|c|c|c|} 
\hline 
\raisebox{0pt}[12pt][6pt]{Parameter} & 
\raisebox{0pt}[12pt][6pt]{Plane Radiator} & 
\raisebox{0pt}[12pt][6pt]{Sawtooth Radiator} \\
\raisebox{0pt}[12pt][6pt]{~~}&
\raisebox{0pt}[12pt][6pt] {($30^{\circ}$)}&
\raisebox{0pt}[12pt][6pt] {($0^{\circ}$)} \\
 \hline
$\sigma _{\gamma}$ & 13.5 mr & 11.8 mr\\
$<N_{\gamma }>$ & 15.5 & 13.5 \\
$\sigma _t$ & 4.5 mr & 4.8 mr \\
$\sigma _t (MC)$& 3.9 mr & 3.8 mr \\
$\sigma _t (CLEO)$ & 4.0 mr & 2.9 - 3.8 mr\\
\hline
\end{tabular}
\end{center}
\end{table}
\vspace*{3pt}
Table 1 shows a summary of the predicted and achieved
values of these variables in the two data sets discussed in this paper and in the corresponding
Monte Carlo simulation, as well as the expectations for the final system. 
Fig.~2 shows the 
measured resolution per track $\sigma _t$ as a function of the 
background subtracted number
of photons detected for the flat radiator and Fig.~2 shows the
corresponding curve for the sawtooth radiator. The curves are 
fit to the parameterization $(a/\sqrt{N_{ph}})^2+b^2$. 

The data shown, although preliminary, show a good understanding of the system and give confidence that
the predicted level of efficiency versus resolution will be achieved in CLEO III. An active analysis
program is under way to study the additional data available. 

\section{Conclusions}
Both the major systems for the CLEO III detectors are well under way. Test runs data support the 
expectations of excellent performance of both the silicon tracker and the Ring Imaging Cherenkov 
detector. This will lead to a quite exciting physics program expected to start in 1999.

\section{Acknowledgements}
The author would like to thank her colleagues in the CLEO III Si3 and RICH groups for their excellent work 
reported in this paper. Especially noteworthy were P. Hopman, H. Kagan, I. Shipsey and M. Zoeller for their help in
collecting the information relative to the Si3 tracker, S. Anderson, S. Kopp, E. Lipeles R. Mountain, S. Schuh, A. Smith,
 T. Skwarnicki, G. Viehhauser
that were instrumental to a successful test beam run. Special thanks are due to C. Bebek and S. Stone for their 
help throughout the length of the RICH project.

\end{document}